\RequirePackage{lineno}	
\documentclass{PoS}
\usepackage{subcaption}
\usepackage{cite}
\usepackage[utf8]{inputenc}
  \usepackage{lineno}
\newcommand{\pT}{$\rm p_{T}$}

\newcommand{\sNN}{$\sqrt {s_{\rm NN}}$~}
\newcommand{\s}{$\sqrt {s}$~}

\newcommand{\DirPho}{$\gamma_{dir}$}
\newcommand{\gammaRich}{$\gamma_{rich}$}
\newcommand{\piZro}{$\pi^{0}$}

\newcommand{\Rjet}{$\rm R_{jet}$}

\newcommand{\ET}{ $\rm E_{T}^{trig} $}
\newcommand{\IAApythia}{$\rm I^{\small{PYTHIA}}_{AA}$}

\title{Semi-inclusive jets recoiling from \DirPho~and \piZro~
 triggers in central Au+Au collisions at \sNN=~200 GeV in STAR}

\ShortTitle{\DirPho+jet and \piZro+jet in STAR}

\author{\speaker{Nihar Ranjan Sahoo (for the STAR Collaboration)}\\
        Texas A\&M University, Texas, USA\\
        E-mail: \email{nihar@rcf.rhic.bnl.gov}}

\abstract{We report the first measurement of fully corrected semi-inclusive charged recoil jets for both direct-photon (\DirPho) and \piZro~triggers, within two trigger transverse energy ranges between 9 GeV and 15 GeV, in central Au+Au collisions at \sNN=~200 GeV using data with an integrated luminosity of 13~$\rm nb^{-1}$ collected by the STAR experiment. A comparison between \piZro-triggered recoil jets in p+p collisions and PYTHIA is discussed. A comparison is also presented between the suppression of \DirPho- and \piZro-triggered recoil jets in central Au+Au collisions with respect to their corresponding PYTHIA references. A strong and similar level of suppression is observed in recoil-jet yields as a function of jet transverse momentum for \DirPho+jet and \piZro+jet.}
\FullConference{International Conference on Hard and Electromagnetic Probes of High-Energy Nuclear Collisions\\
		30 September - 5 October 2018\\
		Aix-Les-Bains, Savoie, France}

\begin{document}
\section{Introduction}
Jet quenching is an important signature of the hot and dense QCD matter produced in heavy-ion collisions known as the Quark-Gluon Plasma (QGP)~\cite{STARpaper}. A direct photon (\DirPho) produced in coincidence with a recoil jet  (\DirPho+jet) is a good probe of the parton energy loss in the QGP~\cite{WangDirPho1,WangDirPho2, WangDirPho3}: since \DirPho~is colorless, it does not experience strong interaction with the medium, and hence the initial energy of the recoil parton (predominantly a quark from QCD Compton scattering) is constrained by the \DirPho~trigger energy. The measurement of \piZro+jet coincidences, in which the recoil parton from a \piZro~trigger can be a quark or gluon, complements the analysis with the \DirPho~triggers. A comparison between the \DirPho+jet and \piZro+jet  distributions in the same kinematic region can provide unique insight into the dependence of parton energy loss on the color factor, the path length, and the initial energy of the parton. 
We present the fully corrected results of this measurement at RHIC, for two trigger \ET~bins between 9 and 15 GeV, and charged-jet radius \Rjet=0.2.
 
\section{Dataset and experimental details}
The data analysed for these proceedings correspond to Au+Au and p+p collisions at \sNN=~200 GeV collected during the years of 2014 and 2009, respectively, at RHIC. 
 	A sample of 0-15\% central Au+Au events is analysed.	
 For this analysis, the STAR Barrel Electromagnetic Calorimeter (BEMC) and a special trigger condition are used to identify \piZro/\gammaRich-triggered events. Here \gammaRich~represents an enriched sample of \DirPho~and with an admixture of photons from \piZro~decays. The purity of direct photons varies within 65\%--85\% for the trigger range of 9 < $\rm E_{T}^{trig} $< 20 GeV. In order to discriminate between \piZro~and \gammaRich~triggers, a Transverse Shower Profile (TSP) method is used~\cite{STARPLBGammaHadron}. The Time Projection Chamber (TPC) is used to measure charged tracks within $|\eta|$<1.0 and 0.2 < \pT < 15 GeV/$c$.

 Charged jet reconstruction is performed using the anti-$k_{\rm T}$ algorithm~\cite{antikt} from the FastJet package~\cite{JetArea}. In this analysis, charged particles with transverse momentum between 0.2 < $p_{T}^{\rm const}$ < 15 GeV/$c$ are included as constituents for jet reconstruction in both Au+Au and p+p collisions. A fiducial cut in jet pseudorapidity, $\rm |\eta_{jet}| <1-$\Rjet, is used to ensure jets fully fit into the TPC acceptance. Semi-inclusive charged recoil jet-\pT~distributions are analysed within two \ET~bins for both \DirPho-~and \piZro-triggered events: 9 < \ET < 11 GeV and 11 < \ET < 15 GeV. 
These narrow trigger \ET~bins provide access to the parton energy loss with relatively narrow conditions for the parton energy. 

\begin{figure}[htbp]
  \centering
 \includegraphics[width=0.4\textwidth]{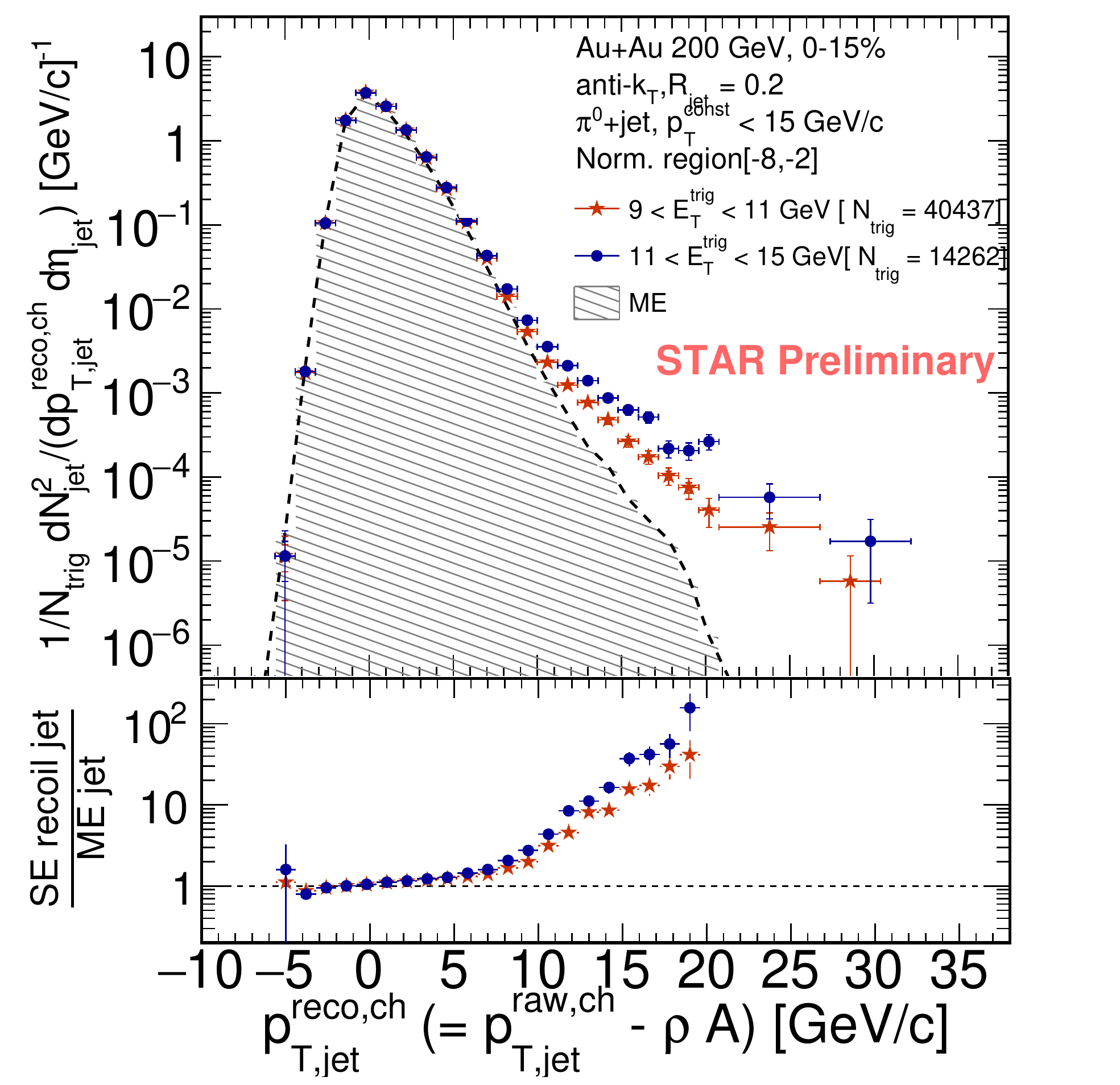}
 \includegraphics[width=0.4\textwidth]{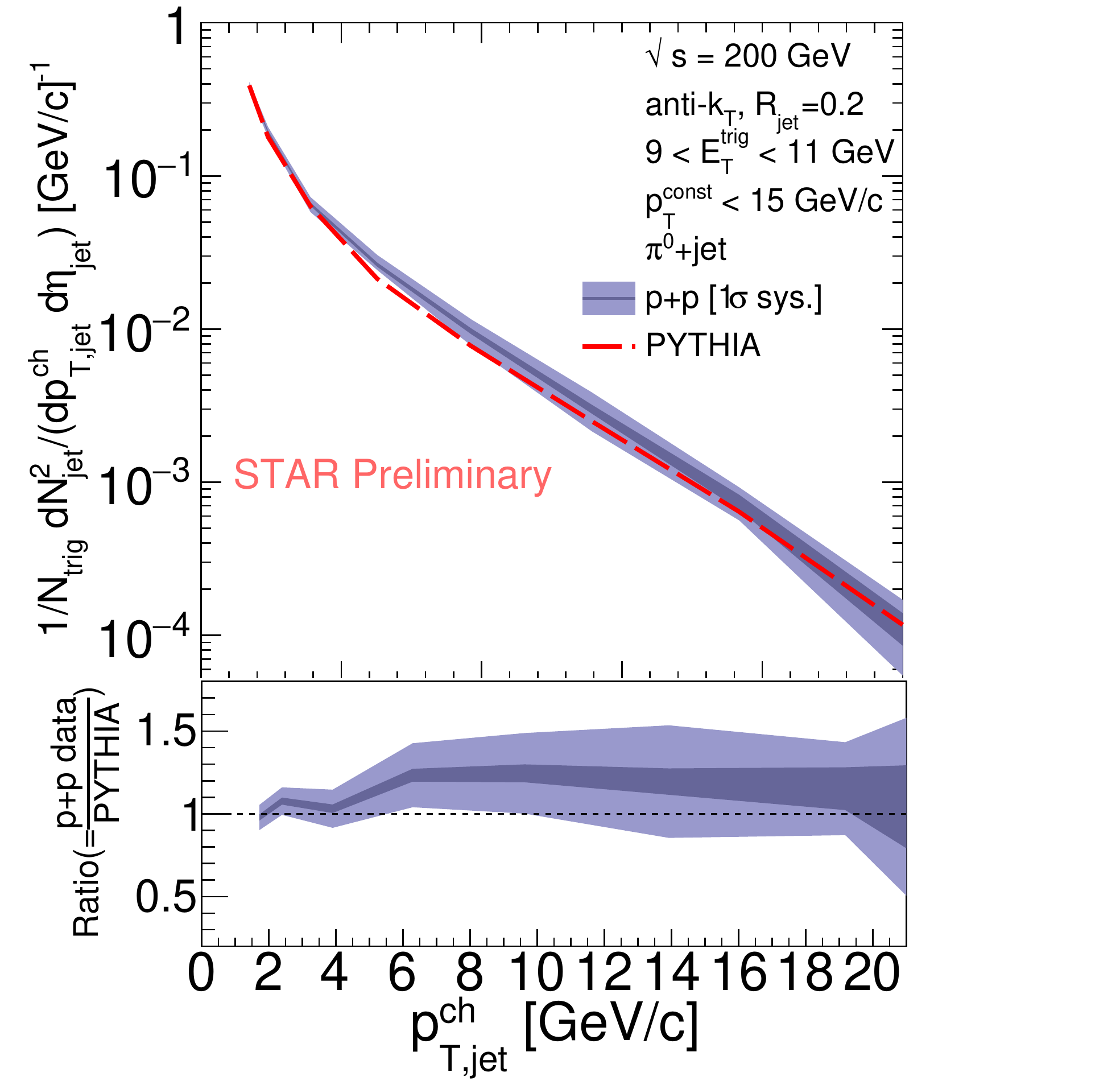}
\caption{Left: Raw semi-inclusive recoil-jet $\rm p_{T,jet}^{reco,ch}$ distributions, from \piZro-triggers, for  trigger \ET~bins: 9 < \ET < 11 GeV (stars) and 11 < \ET < 15 GeV (circles). The ME distribution is shown as hatched lines. Error bars represent statistical uncertainties only. The lower panel shows the ratio of SE to ME distributions for the two \ET~bins. Right: Fully corrected recoil-jet \pT~distributions from \piZro-trigger for 9 < \ET < 11 GeV in p+p collisions; lighter and darker blue bands represent ($1\sigma$) systematic and statistical uncertainties, respectively. The dashed line represents the PYTHIA8 expectation. The ratio of STAR p+p data to PYTHIA8 is shown in the lower panel.}  
  \label{RecoilJet}
\end{figure}

The recoil-jet region is defined as $\Delta\phi \in [\pi-\pi/4, \pi+\pi/4]$, where $\Delta\phi$ is the azimuthal angle between the trigger tower and the jet-axis.
The left panel in Fig.~\ref{RecoilJet} shows the distribution of recoil jets with \Rjet=0.2 as a function of $\rm p_{T,jet}^{reco,ch}$ (= $\rm p_{T,jet}^{raw,ch} - \rho A$) for \piZro~triggers in the two \ET~ranges, where  A represents the jet area with radius \Rjet, and $\rho$ is an estimation of the uncorrelated energy density, calculated event-wise~\cite{JetArea}. Uncorrelated background yield is determined using an event-mixing technique (ME) as was done in the STAR hadron+jet analysis~\cite{STARhjet}. The distributions for real events are denoted ``SE" for "same event". The distribution of $\rho$ for the ME and SE populations are the same. A clear trigger \ET~dependence is observed in recoil-jet $\rm p_{T,jet}^{reco,ch}$ spectra. The SE recoil jet $\rm p_{T,jet}^{reco,ch}$ distribution dominates over the ME jet $\rm p_{T,jet}^{reco,ch}$ distribution above the trigger-\ET~range, as shown in the lower left panel of Fig.~\ref{RecoilJet}. 

This analysis involves several steps to obtain the fully corrected \piZro+jet and \DirPho+jet \pT~spectra: i) discrimination between \piZro~and \gammaRich~triggered events using the TSP method; ii) reconstruction of charged recoil jets in SE; iii) determination of the uncorrelated background yield using the ME population; iv) subtraction of ME from SE recoil distributions; iv) determination of the response matrix due to detector effects and \pT-smearing; v) unfolding the ME subtracted recoil jet distribution; vi) conversion from \gammaRich+jet to \DirPho+jet using the estimated percentage of purity (as previously done in ~\cite{STARPLBGammaHadron}), and vii) an estimation of systematic uncertainties by varying unfolding methods, priors, ME normalisation region, and \DirPho~purity.

For \piZro+jet and \DirPho+jet distributions in p+p collisions, PYTHIA events are simulated using PYTHIA8~\cite{pythia} with the default setting, at \s=~200 GeV, to compare with the data. For unfolding p+p data, a full GEANT simulation is performed to reconstruct the response matrix. On the other hand, a fast simulation is used to reconstruct the response matrix for both the detector and uncorrelated background fluctuation effects in central Au+Au collisions as was done in~\cite{STARhjet}.


\begin{figure}[htbp]
  \centering
      \includegraphics[width=0.37\textwidth]{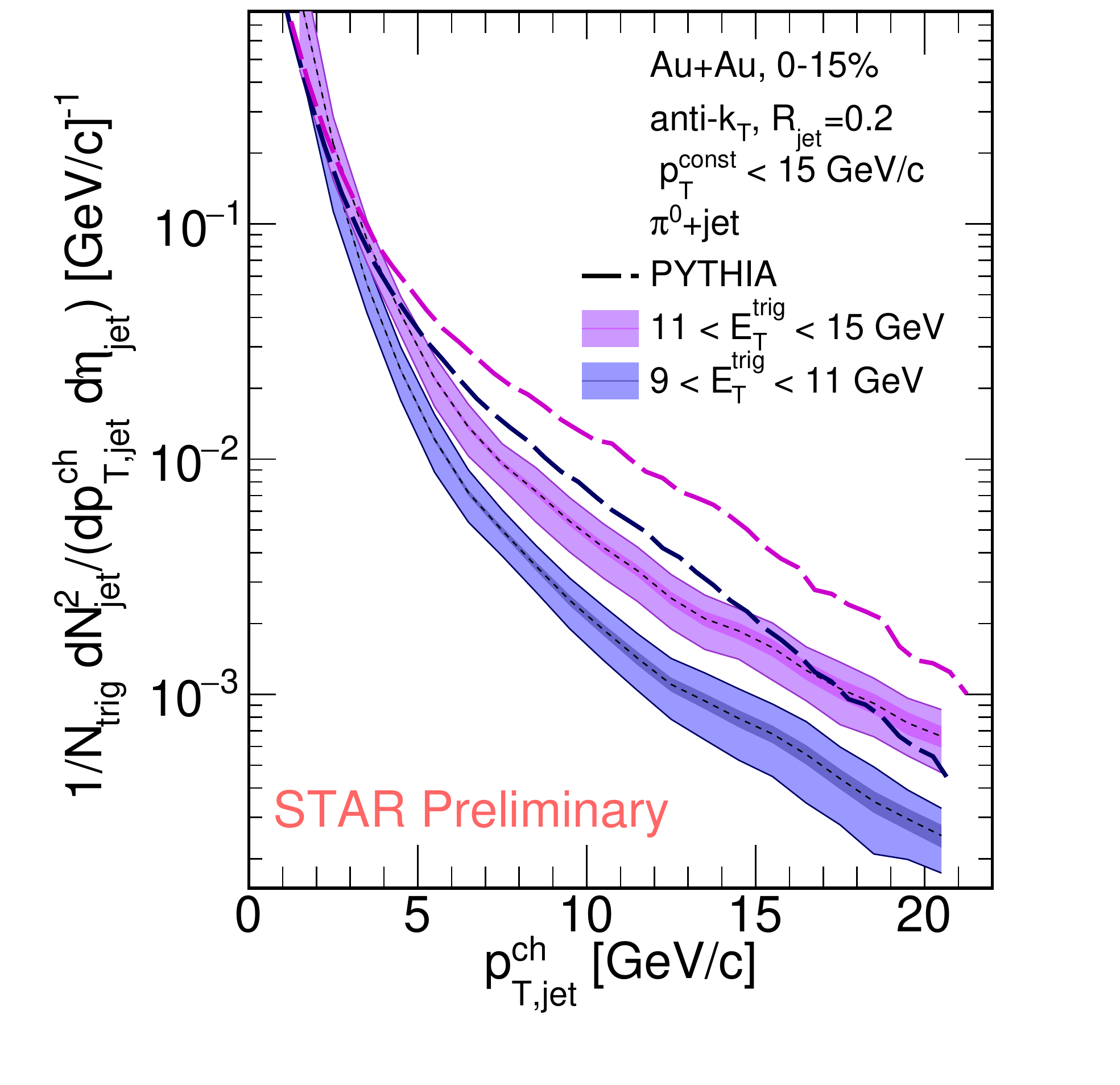}  	  	 	
     \includegraphics[width=0.37\textwidth]{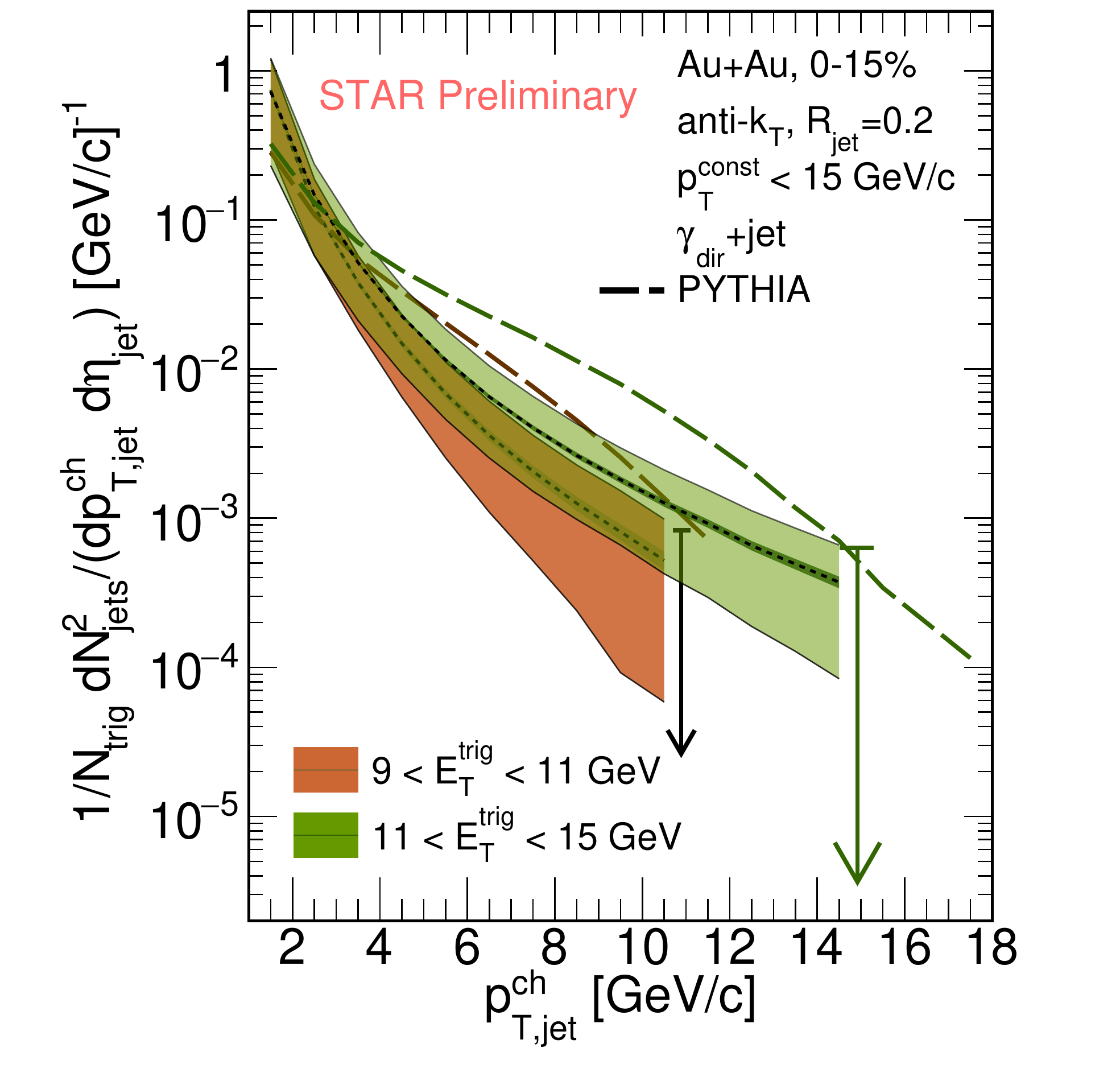}  	  	 

\caption{Left: The \piZro~triggered fully corrected recoil jet \pT~($\rm p_{T,jet}^{ch}$) distributions for two trigger \ET bins: 9 < \ET < 11 GeV (blue band) and 11 < \ET < 15 GeV (magenta band). Right: Same for the direct-photon trigger with 9 < \ET < 11 GeV (orange band) and 11 < \ET < 15 GeV (green band). Lighter and darker bands represent systematic and statistical uncertainties, respectively. Downward arrows represent upper limit in the yield. Dashed lines represent respective PYTHIA expectations in  each case.}  
  \label{CorrRecoilJetpT}
\end{figure}


\begin{figure}[htbp]
  \centering
    \includegraphics[width=0.5\textwidth]{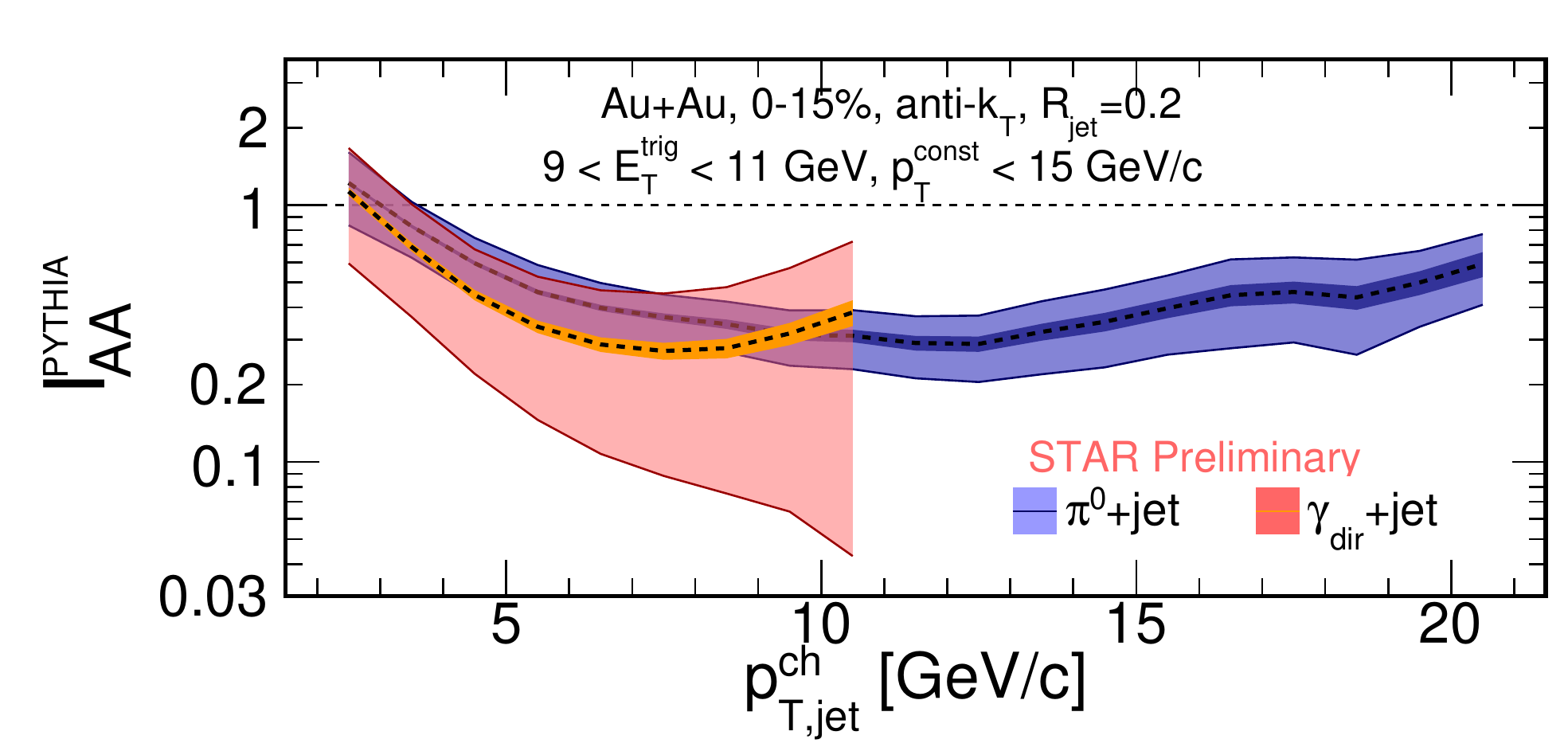}
  \includegraphics[width=0.5\textwidth]{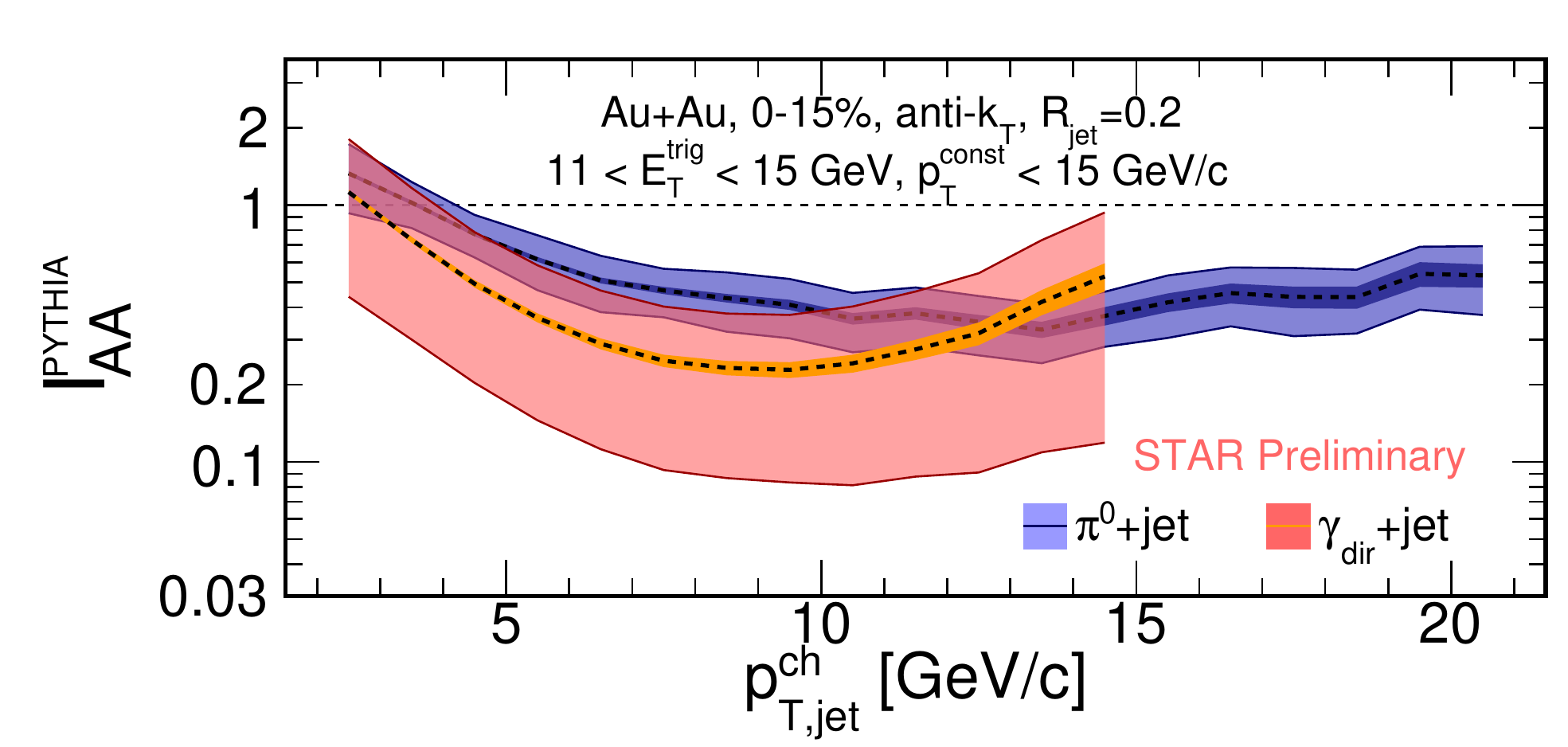}

\caption{\IAApythia~vs.  $\rm p_{T,jet}^{ch}$ for \DirPho- (red band) and \piZro-trigger (blue band) recoil charged jet. Top: 9 < \ET < 11 GeV. Bottom: 11 < \ET < 15 GeV. Lighter and darker bands represent systematic and statistical uncertainties, respectively. }
  \label{iaaplot}
\end{figure}

 
 \vspace{-15pt}
 
 \section{Results and discussion}

Figure~\ref{RecoilJet} (right) shows the fully corrected charged recoil jet $\rm p_{T,jet}^{ch}$ distribution triggered by \piZro, with \Rjet = 0.2 and 9 < \ET < 11 GeV in p+p collisions at \s~= 200 GeV. The PYTHIA distribution is in agreement with the measured p+p distribution within uncertainties, providing validation of PYTHIA for calculating the p+p baseline in this analysis.

Figure~\ref{CorrRecoilJetpT} shows the fully corrected charged recoil jet $\rm p_{T,jet}^{ch}$ distribution triggered by \piZro~(left plot) and \DirPho~(right plot) for 9 < \ET < 11 GeV and 11 < \ET < 15 GeV, with \Rjet=0.2 in 0-15\% central Au+Au collisions. For \piZro+jet, the dominant contribution to the systematic uncertainties is due to the unfolding procedure, whereas for \DirPho+jet, the dominant contributions include both the unfolding procedure and the determination of the  \DirPho-purity. A clear \ET~dependence of the recoil-jet $\rm p_{T,jet}^{ch}$ spectrum can be observed for \piZro+jet; a similar indication can also be seen for \DirPho+jet within the systematic uncertainties. 

We quantify the suppression of the recoil-jet yield in $\rm p_{T,jet}^{ch}$ via $\rm I^{PYTHIA}_{AA}( \rm p_{T,jet}^{ch})$ = $\frac{\rm Y^{Au+Au}( \rm p_{T,jet}^{ch})} {\rm Y^{PYTHIA}( \rm p_{T,jet}^{ch})}$, where $\rm Y^{Au+Au}( \rm p_{T,jet}^{ch})$ and $\rm Y^{PYTHIA}( \rm p_{T,jet}^{ch})$ represent the per-trigger recoil-jet yields in central Au+Au and PYTHIA collisions, respectively. Figure~\ref{iaaplot} compares \IAApythia ($\rm p_{T,jet}^{ch}$)~for \piZro+jet and \DirPho+jet for two \ET~trigger bins: 9 < \ET < 11 GeV and 11 < \ET < 15 GeV. A strong suppression of the \piZro+jet and \DirPho+jet yields is observed for both trigger \ET~bins with a similar level of suppression for the two trigger types. Comparison of  \piZro+jet suppression, for 9 < \ET < 11 GeV with previously published hadron+jet results for a wider trigger \pT~range 9 < ${\rm p_{T}^{trig}}$ < 30 GeV/$c$, shows that \piZro+jet and hadron+jet have a similar level suppression---roughly 60\%-70\% between 10 < $\rm p_{T,jet}^{ch}$ < 20 GeV/$c$. \IAApythia($\rm p_{T,jet}^{ch}$)  of \piZro+jet is compared for two different trigger \ET~and shows no clear trigger \ET~dependence for \Rjet=0.2.

\vspace{-7pt}
\section{Summary and outlooks}	
We present the first measurement of fully corrected \DirPho+jet and \piZro+jet distributions in p+p and central Au+Au collisions at \sNN=~200 GeV for charged-particle jets with \Rjet=0.2 in the range $\rm p_{T,jet}^{ch}$< 20 GeV/$c$. It is found that in p+p collisions, the \piZro+jet $\rm p_{T,jet}^{ch}$ distribution is in agreement with PYTHIA and the PYTHIA results are used as references for the corresponding Au+Au results. In central Au+Au collisions, strong suppression of \DirPho+jet and \piZro+jet is observed over the range 9 < \ET < 15 GeV for \Rjet=0.2. This suppression is the same for both trigger types.  More detailed analyses are ongoing to investigate the redistribution of lost jet energy at a larger jet radius (\Rjet=0.5) and higher \ET ( > 15 GeV).

\vspace{-10pt}
\section*{Acknowledgments}
\vspace{-8pt}
  
This work was supported by the US DOE under the grant DE-FG02-07ER41485. 

 \vspace{-10pt}


\end{document}